# Simultaneous Crystallization Effects in Multiple Levitated Plasma-Functionalized Graphene Nanoflake Nanofluid Droplets

*Adam McElligott, André Guerra, Alejandro D. Rey, Phillip Servio\**

Department of Chemical Engineering, McGill University, Montreal, Quebec H3A 0C5, Canada

*phillip.servio@mcgill.ca

ABSTRACT

   Acoustic levitation is a container-free method for examining novel crystallization effects, though liquid-to-solid phase change has seen little investigation for levitated nanofluids. Recent developments have allowed for examining the morphological and temperature evolution of multiple levitated nanofluid droplets freezing simultaneously. The fundamental effect of adding nanoparticles to a levitated crystallization system is crystal growth rate enhancement from improved mass transfer at the growing solid front. Nucleation times are unaffected as freezing is initiated by secondary ice nucleation particles (INPs). Instead, the enhancement produces higher instantaneous nucleation pressures and more cracking in the primary ice shell. In turn, more INPs are ejected, resulting in faster protrusion formation on the droplet surface (hastened further in systems containing adjacent droplets). The crystal matrix also includes more defects, resulting in liquid escaping and forming beads at the droplet base and optical clarity loss. During crystal decomposition, thermal gradients create convective currents dampened by the same transport phenomena that enhance crystal growth. Suspension loss after a crystallization-decomposition cycle reduced opacity and light absorbance such that the droplets were 62% closer in appearance to water. However, the non-isobaric, sample-encompassing cooling process resulted in smaller particle clusters than if the droplets were frozen on a solid surface.

1. INTRODUCTION

Levitation techniques have garnered increasing interest in laboratory studies, particularly those in engineering and biomedicine using liquid or partially liquid samples.[1] These devices' advantages include eliminating container effects, a lower risk of specimen contamination, and high degrees of environmental and morphological control.[1-3] Moreover, the samples have no specific chemical, physical, or compositional requirements, as there is no interaction with surfaces that may modify interfacial energy or add wetting as part of the experimental conditions.[4,5] Therefore, various applications have emerged in the study of levitated liquids, such as novel mixing or drying methods and new ways to investigate liquid properties.[4,6-8]

Levitation can be achieved through optical or electrodynamic means or by simply placing a droplet into a laminar air stream, called "free-fall" levitation.[9-11] Acoustic levitation has also become a prominent technique in which a transducer generates standing waves that bounce off a concave reflector and trap particles at uniaxial nodes.[12] It has seen notable use in crystallization studies, where nucleation and freezing have been examined in water, mimicking atmospheric ice formation, and in more complex compounds like clathrate hydrates to eliminate the effects of impurities, grain boundaries, or other system-specific defects.[6,7,13] These studies have found distinct liquid-to-solid phase transition behaviors compared to more common, heterogeneously nucleated processes and have indirectly provided insight into the effects of vessel walls on crystallization. For instance, levitated droplets immediately form a solid shell around the entire sample upon nucleation.[14,15] A significant interfacial pressure develops at the new liquid-solid interface, creating defects and releasing secondary solid particles.[15] The strength of this shell determines the number of secondary particles that, when released, initiate the nucleation of nearby droplets: an effect rarely seen when crystallization begins from a solid boundary.[6,7]

Levitated nanofluids (i.e., levitated suspensions of dispersed nanoparticles) have received little investigation regarding ice nucleation, crystal growth, and the morphology of frozen masses. Instead, the focus has been on evaporation processes for chemical synthesis or particle self-assembly.[1, 4, 16] For example, Raju et al. (2019) irradiated a colloidal polystyrene droplet with a laser to assemble millimetric discs with photonic crystal behavior upon evaporation.[16] These laser-based processes take advantage of an open configuration, so researchers have direct access to their samples and can use cameras to observe the changing morphology of the liquid droplet or modify the laser direction.[1, 3, 17-19] Levitated nanofluid crystallization on the other hand, has required the system to be completely contained to achieve freezing.[2, 20, 21] Droplet sizes are small enough that cooling from room-temperature air is insignificant compared to laser irradiation, whereas cooling temperature control is much more difficult in open systems.[7] Another obstacle has been that no proposed direct application for freezing levitated nanofluids exists beyond general scientific interest. However, such investigations can indirectly provide insight into the effects present in existing applications using solid surfaces by examining crystallization without those surfaces.

This is not to say that cooling or crystallization studies have never been completed for levitated nanofluids. Yudong et al. (2015) examined the supercooling degree and nucleation behavior in graphene oxide nanofluids. They found that the supercooling degree before nucleation may have decreased in the presence of nanoparticles and that more nucleation sites may have also been created.[21] Supercooled, levitated graphene nanofluid surface tension has also been investigated using physical methods and artificial neural networks.[5, 22] Nevertheless, these studies have the same closed-system limitations as in other levitated crystallization experiments. Namely, there was no direct visual observation of droplet morphology nor temperature readings from an infrared (IR) camera, meaning that a thermocouple had to be inserted into the droplet, and a solid interface for nucleation was introduced. In addition, these studies focus on the nucleation or pre-nucleation states of single droplets and do not examine total crystallization morphology or melting behavior. However, recent developments in the field now allow for freezing liquid droplets suspended in an open acoustic levitator using a cryogenic stream.[7, 23] The TinyLev, developed by Marzo et al., uses multiple ultrasonic transducers to levitate several droplets on a single axis.[12] Instead of a cooling chamber, a custom cryogun can be used to freeze these droplets simultaneously.[6, 7] There is no obstructing interface, so direct and clear images from digital or IR cameras can be captured, and droplet size and nodal location can be precisely controlled. Therefore, bridging the gap between the levitated nanofluid heating (open system) and cooling (crystallization) systems is now possible.

This study examines the morphological and thermal behavior of aqueous 100 ppm plasma-functionalized graphene nanoflake (GNF) droplets suspended in an acoustic field during crystallization. Plasma-functionalized graphene does not require a surfactant to be fully dispersed in solution, so only water and the nanoparticle will be present.[24] Moreover, many similar studies use an analogous particle, graphene oxide, so they can be used as a strong point of comparison.[2, 20, 21, 25] Up to three droplets will be frozen simultaneously, and two coincident cameras will capture any interdroplet effects directly. To our knowledge, this is the first time completely free-floating nanofluid crystallization will be observed, in addition to the effects of concurrent crystallization events. Furthermore, this study will go beyond the nucleation and initial freezing stages and examine solidification (bulk crystal growth) and melting (crystal dissolution). Dominant nucleation factors and mechanisms, interfacial and bulk transport phenomena, and colloidal effects during the freeze-thaw cycle will be investigated. Therefore, this study will focus on the system's complex interactions between geometry, phase transition, suspension dynamics, and morphology during crystal growth and decomposition.

2. MATERIALS AND METHODS

2.1 Experimental Equipment and Setup

The following section details the experimental setup for levitated nanofluid crystallization. It combines two previous works: the TinyLev acoustic levitator from Marzo et al. and the cryogun modification for freezing developed by McElligott et al., both of which have been used successfully in several other studies.[6, 7, 23, 26] While critical features will be outlined here, more information can be found in these sources. A simplified version of the experimental setup is shown in **Figure 1**.

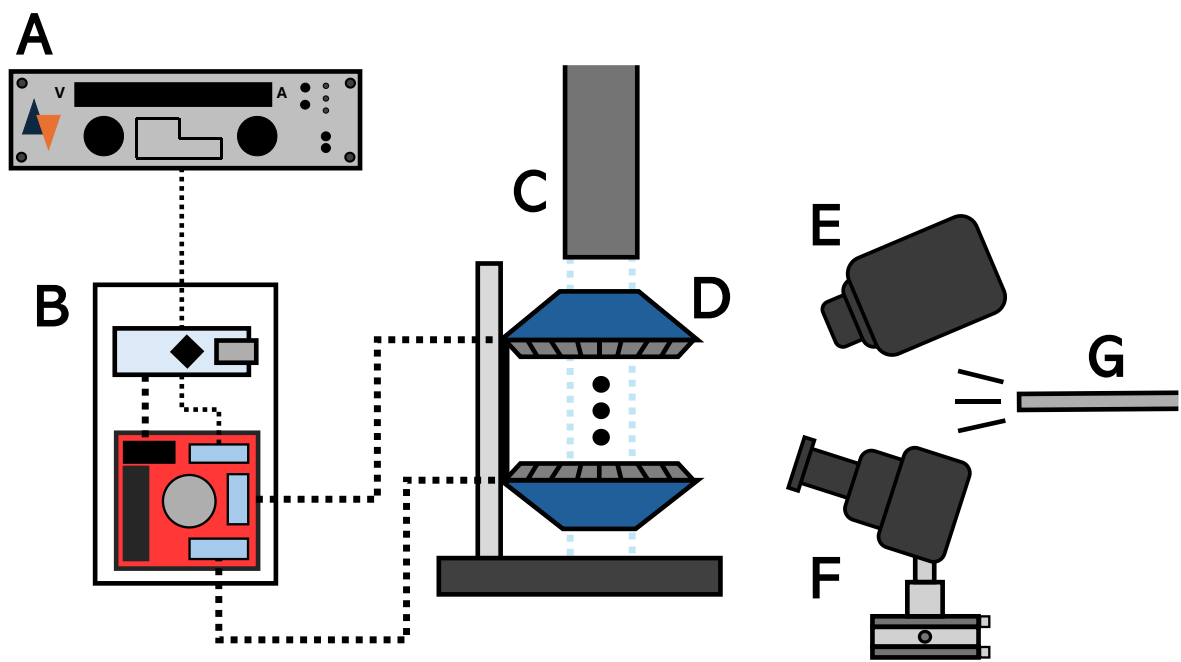

**Figure 1.** Schematic of the experimental setup (simplified). The schematic includes the following: a power supply (A), a driving board (B), a custom-built cryogun emitting a cryogenic stream (C), the TinyLev acoustic levitator (D), an infrared camera (E), a digital camera on a multi-axis

positioner with its lens (F), and an LED light source (G). The camera orientations are only for clarity; they have perpendicular lines of sight and are level with the droplets during operation.

The TinyLev is an open, single-axis acoustic levitation device shown as (D) in **Figure 1**. It consists of two acoustic arrays spaced 7.5 centimeters apart. Each array contains 36 acoustic transducers arranged into three rings and has an opening directly in its center that is vertically aligned to the axis of levitation. The standing wave created by this configuration produces 11 potential nodes for placing liquid droplets if operating at 40 kHz, though previous studies have only achieved the simultaneous freezing of five droplets.[7] However, if only the three most central nodes are occupied, the cameras can be placed closer to the levitated samples, resulting in clearer images for analysis. Moreover, the central nodes are the most stable for crystallization, so experimental repeatability is also improved. Therefore, only those positions were used and numbered (from top to bottom) 1 to 3. The power supply, denoted as (A), is a Delta Elektronika SM 70-AR-24 operating at 0.8 A and 10 V. Powering the TinyLev with these settings creates the identical acoustic field to previous free-floating pure water and tetrahydrofuran hydrate crystallization studies, so that these can be used for comparison.[6, 7] Rounding out the levitation portion is the driving board (B), consisting of an Arduino Nano and an L297N Dual H-Bridge motor driver, through which the generated power runs. The driving board produces and then amplifies square wave excitation signals that are subsequently sent to each array. As the system is open, conditions in the laboratory must be reported: the average room temperature was 22.2 ± 0.4 ºC, and the average humidity was 22.5 ± 1.6 %, consistent with previous studies. Note that the humidity was maintained through each experimental run as the air supply came directly from the room.

A cylindrical, custom-built cryogun made from stainless-steel 316 (C in **Figure 1**) was placed directly above the topmost array to freeze the droplets. From the base of the cylinder, which was centered on the axis of levitation, the distances from positions 1, 2, and 3 were 6.4, 6.9, and 7.4 cm, respectively. When the cylinder is filled with liquid nitrogen, it cools, and the base temperature stabilizes to -55.5 ºC. The air temperature around the base also decreases, increasing air density. This creates a significant cooling stream when the air begins to fall due to gravity, a stream maintained during the entire experimental run sufficient to freeze liquid samples at all positions. IR imaging was done using a Jenoptik IR-TCM 384 infrared camera (E). It was calibrated to a measurement accuracy of 1 to 2 % in the −20 to 20 ºC temperature range and had a NETD temperature resolution of less than 0.08 ºC. The thermal images produced by this camera were used to determine temperature conditions during each run. Digital grayscale images were taken using an 18.0-megapixel Canon EOS 60D DSLR camera with an MP-E 65 mm f/2.8 1-5× macro lens. It is marked as (F) in **Figure 1**, along with the OptoSigma multi-axis manual translation stage on which it is mounted. To optimize digital imaging, a black backing was inserted into the TinyLev behind the droplets, and a fiber optic LED light (denoted as G) was added above the camera, pointing down at the droplet(s). These additions increase the contrast of the droplet and the background, making the liquid-air boundary more clearly distinguishable. The IR and digital cameras were perpendicular to each other and had lines of sight that were parallel to the table (normal to the axis of levitation). Therefore, they would capture different "sides" of the droplet at any instant, though both images match almost exactly due to the droplet spinning in the acoustic field, a known phenomenon.[7]

## 2.2 Experimental Procedure and Data Acquisition

The acoustic field in the TinyLev was initiated (by turning on the power supply) before data acquisition would commence. Using a 3 mL syringe, the droplets could be placed directly into the three nodal positions. Seven configurations were investigated: the individual droplets in three different nodes in the field and the multi-droplet combinations in positions 1 and 2 (1/2), 1 and 3 (1/3), 2 and 3 (2/3), and 1, 2, and 3 (1/2/3). Plasma-functionalized GNFs dispersed in reverse osmosis (RO) water at a 100 ppm concentration were chosen because (1) no surfactant was required to maintain the suspension, (2) the concentration was similar to those in other studies, and (3) the initial droplet was very dark such that dispersion losses or internal flows would be more evident. The GNFs were produced, characterized, and dispersed in McGill University's Plasma Processing Laboratory before this study began. Therefore, only a few critical details will be provided here, and further information, including the functionalization process and imaging of the GNFs, is available in Legrand et al. (2016).[24] The GNFs were created by taking a pure powder of graphene sheets and exposing them to an argon plasma under an air atmosphere, adding hydrophilic oxygen functional groups to their surface. These GNFs are 10 atomic layer-thick rectangular prisms with $100 \times 100$ nm$^2$ planar dimensions. Their surface atomic composition is 14.2% oxygen, with the remainder mostly being carbon. Though ultrasonication is initially required to disperse the particles, with these characteristics, the nanofluid can remain stable in the aqueous form for years without agglomeration or settling.[24]

The cryogun was filled with liquid nitrogen, away from the setup, to start an experimental run. One minute would be allowed to elapse so that the cryogun surfaces were adequately cooled; data capture was then initiated from both the IR and digital cameras. Following this, the cryogun was placed above the opening of the TinyLev's topmost array. It emitted a stream of cold air sufficient to freeze all present liquid droplet(s) regardless of position or configuration. The cooling period was set to two minutes, ensuring the samples' complete solidification. When this period ended, the cryogun (i.e., the cooling stream) was removed from the setup. Data capture would continue for one minute while the droplets melted. In addition to the thermodynamic and morphological data captured by the two cameras, one post-processing step was required to quantify nanoparticle dispersion loss. ImageJ software was used to determine the mean gray value (MGV) of the initial and final liquid droplets. The MGV can be described as how close the average color of the sample is to black (which has a value of 0) or white (which has a value of 255) in grayscale. In our case, a droplet starting at a low MGV but ending at a high one after a crystallization-decomposition cycle would indicate a loss of nanoparticle dispersion. It is important to note that this value cannot be used to obtain precise information about the system's colloidal parameters (e.g., aggregate sizes or effective concentrations). Additionally, post-experimental physical handling of the samples could potentially improve the dispersion, so such parameters cannot be obtained accurately. Instead, the MGV provides a relative change in dispersion that can be compared between the various configurations in this study. A total of 15 replicates were completed for each individual nodal position and multi-droplet configuration. This makes 105 replicates examined and analyzed, all of which were over a single freeze-thaw cycle.

3. RESULTS AND DISCUSSION

3.1 Nucleation of Levitated Graphene Nanofluid Droplets

Aqueous 100 ppm GNF nanofluid droplets were suspended and frozen to form solid crystal masses in an acoustic field using a cryogenic stream. A crystallization system could contain a single droplet or multiple (up to three) droplets in a particular configuration. **Figure 2** shows an example of the initial stages of the crystallization process in the position 2 system through both morphological and thermal behavior. Both behaviors exhibited in these stages are characteristic of levitated liquid freezing: regardless of their position or the number of droplets present, the droplets displayed the same fundamental properties.[6, 7] Before the cryogun is placed and the cooling stream acts on the droplet (A), the droplet can be considered at rest in the acoustic field. It adopts an oblate spheroid shape and moves very little. When the cryogun is added to the system (B), the droplet adopts a lower eccentricity shape (i.e., circularizes). Note that while circularization is marked as a time of 1 second in **Figure 2**, this moment is marked as 0 seconds for later time measurements. The cryogenic stream is not clearly evident in the digital images, though the background of the thermal images shows a darker shade of pink, demonstrating its presence. Moreover, the droplets display multiple lower-wavelength colors in the IR image, which shows that they are beginning to cool. After a short time (C), ice nucleation occurs, indicated by two primary behaviors: (1) the immediate formation of a shell around the droplets, as described in the Introduction, and (2) a rise in droplet temperature from the exothermic crystallization event. As the droplet freezes and the solid front grows (D), the droplet continues to cool and some water escapes the shell to form a bead at the droplet base that can be seen in both image types.

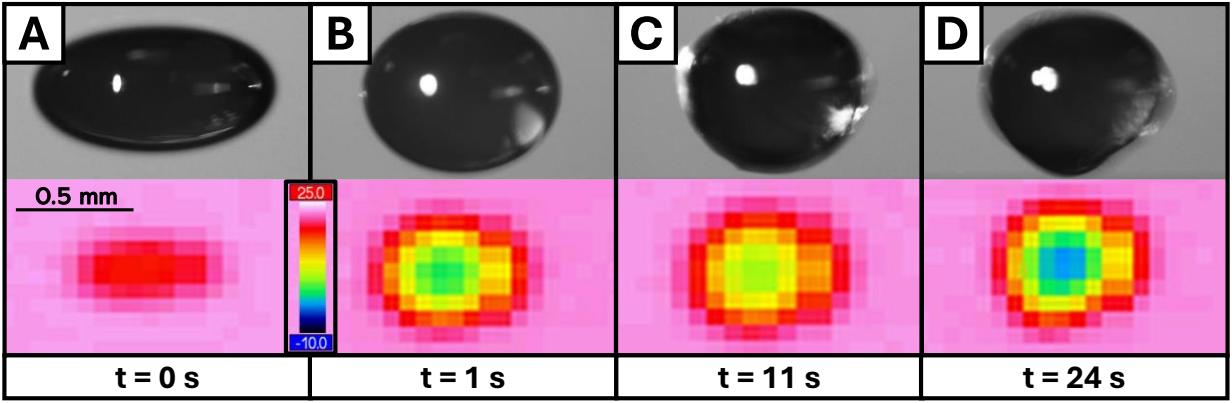

**Figure 2.** An example of the morphological and thermal evolution from initial nanofluid droplet conditions (A) to nucleation (C) and bead formation (D) at position 2. There is no cooling stream at 0 seconds. Then, at 1 second, the cryogun is added so the droplets circularize and cools (B).

### 3.1.1 Initial Droplet Conditions

In **Figure 2**A, the free-floating nanofluid droplets are not perfectly spherical but instead take on the shape of oblate spheroids. In other words, the lateral axis is longer than the vertical one. This characteristic shape arises from a combination of forces present in the acoustic field and within the droplet. Gravity acts downward on the droplet, while the lower array of the TinyLev emits a counteracting acoustic pressure, keeping it stably in the air. At the same time, the upper array emits a weaker acoustic pressure that maintains the droplet position and stops it from moving out of the acoustic region.[23] Together, these forces create acoustic streamlines, called Rayleigh streaming, that form at the air-liquid interface, the boundary layer just before the droplet surface.[1] In turn, the streamlines determine the boundaries of the droplet by forming vortices at its upper and lower halves.[27] Within the droplet, hydrostatic forces and surface tension balance these outer normal forces, giving rise to the droplet shape.[28] Note that some streaming (Eckart streaming) can also occur within the droplet due to viscous damping from the acoustic field, though this will be discussed more in the next section.[1] Additionally, a small amount of convective cooling at the surface caused by acoustic streaming-driven flow likely results in initial droplet temperatures lower than room temperature: 18.8 ± 0.6 ºC, 18.3 ± 0.1 ºC, and 18.0 ± 0.3 ºC for position 1, position 2, and position 3, respectively, compared to 22.2 ºC. This is a greater deviation between the surface and environmental temperatures than what is reported in previous water-based levitation studies.[6,7] However, crystallization studies using GNFs have found improved mass transfer at the solid-liquid interface.[29,30] The same effect may be present where GNFs improve mixing between the colder water at the surface and warmer water within the droplet (at the air-liquid interface), enhancing acoustic stream effects and reducing overall droplet temperature. This was likely exclusively a surface effect, as a temperature difference across the droplet could not be detected. If that had been the case, Marangoni flows could have been induced and the nanoparticles would migrate towards the edge of the droplet.[19]

In addition, the average droplet initial volume at position 1 was 2.24 ± 0.16 µL, and that at position 2 was 2.05 ± 0.14 µL. While these are statistically similar, the volume at position 3 was much greater at 2.69 ± 0.28 µL. To counteract gravity, the acoustic pressure generated by the lower array in the TinyLev is higher than that of the upper array. As the droplet in position 3 is closest to the lower array, a stronger acoustic force acts upon it compared to the other droplets, and it can remain stable at a greater volume without fragmenting.[7] More volume also means more surface area, so there may be more acoustic streaming, which results in a colder surface temperature. While the average initial temperature value is lower at position 3, it is not statistically different from the values at the other positions, indicating that this would only be a minor effect. Finally, the initial volumes are not dissimilar to pure water, likely because the 100 ppm concentration is insufficient to alter surface tension or hydrostatic forces substantially.[7] To check, a Krüss Drop Shape Analyzer

DSA30 was used to determine that the surface tension of a 100 ppm GNF droplet (at 20 ºC) was 69.61 ± 1.75 mN/m compared to 72.04 ± mN/m for pure water, only a 3.37 % difference.

3.1.2 Nucleation of Individual Droplets

Regardless of position or configuration, when the cryogenic stream was added and surrounded the droplets, they transitioned from their initial oblate spheroid form to a more circular one like that in **Figure 2**B. This shape, which persisted as the droplets cooled, was likely caused by the additional downward force that reduced the effects of acoustic streaming such that the balance of normal forces was more equal. In other words, the length values on the lateral and vertical axes became more similar because the primary upward and downward forces on the droplet became closer. **Figure 2**C shows the initial dendritic shell formation and temperature rise, which marked nucleation. Both phenomena have been widely reported in previous levitated crystallization investigations, and the liquid-to-solid phase transition is known to be exothermic.[15] Also, methods that use streams to cool levitated droplets will induce droplet spinning during the cooldown period, resulting in a more uniform transfer of latent heat to the environment across the droplet surface.[11] In turn, this avoids heat accumulation at local maxima such that the primary ice shell has no significant weak spots. The undercooling of the droplet, also called the nucleation temperature, is presented in **Table 1**. These values were similar at all positions in all configurations, between -1 and -2 ºC, likely because there is little variation in cryogenic stream temperature in the 1 cm span across the nodal positions (i.e., similar cooling rates). Critically, there is no significant variation between these values and those for pure water, indicating that adding GNFs had no impact on undercooling.[7]

**Table 1.** The nucleation temperature and nucleation time for 100 ppm GNF droplets at each single position and multi-droplet configuration. The numbers in brackets following these values represent the 95% confidence interval (±).

| Configuration | Single | 1/2 | 1/3 | 2/3 | 1/2/3 |
|---|---|---|---|---|---|
| Nucleation Temperature (°C) | | | | | |
| Pos. 1 | -1.33 (0.86) | -1.83 (0.94) | -1.33 (0.86) | - | -1.33 (0.69) |
| Pos. 2 | -1.76 (0.83) | -1.33 (0.69) | - | -1.50 (0.68) | -1.33 (0.69) |
| Pos. 3 | -1.67 (0.97) | - | -1.83 (1.07) | -1.00 (0.68) | -1.83 (0.79) |
| Nucleation Time (s) | | | | | |
| Pos. 1 | 9.00 (0.81) | 8.80 (0.79) | 8.40 (0.39) | - | 8.60 (0.72) |
| Pos. 2 | 11.60 (1.01) | 10.33 (1.84) | - | 11.27 (0.94) | 8.73 (0.85) |
| Pos. 3 | 11.40 (0.53) | - | 11.33 (0.85) | 11.07 (0.96) | 9.27 (0.89) |

It is important to note that this result does not contradict previous reports that found a substantial reduction in undercooling. This primarily concerns the nucleation mechanism, for which the nucleation time (found in **Table 1**) is a strong indicator. Homogeneous or bulk nucleation is unlikely in this system, not only because more thermodynamically favorable mechanisms are present but because this type of nucleation solely depends on the level of undercooling.[31] It follows that if all droplets have similar undercooling, they should nucleate within similar timeframes. Instead, the position 1 droplet nucleates first, and the position 2 and position 3 droplets nucleate a statistically significant longer amount of time later. Therefore, nucleation is not homogeneous in this system, even though the droplets are free-floating. Heterogeneous nucleation at the air-liquid interface may be the dominant mechanism. Previous studies have suggested that levitated ice formation occurs on the droplet surface in at least 90% of cases.[31] Specifically, as ultrasonic waves cause cavitation by creating concentrations of acoustic pressure at the surface, microbubbles could become entrained and increase in number near the interface and become nucleation sites.[32] This pseudo-heterogeneous nucleation would be more thermodynamically favorable than homogeneous or "true" heterogeneous nucleation as it would require less work. However, the critical parameter for this nucleation type is the surface area: droplets with larger surfaces would have more numerous nucleation sites and a greater probability of nucleating first. In that case, the droplet at position 3 should nucleate first, but again, it nucleates with and after the smaller position 2 and position 1 droplets, respectively. Consequently, this is probably not the dominant mechanism either. Therefore, a different, non-primary nucleation mechanism is likely being explored here, and the undercooling values are not comparable.

Position 1 droplets may nucleate faster due to secondary nucleation from subvisible, aerosolized crystals called ice nucleation particles (INPs) that initiate levitated ice formation in relatively high-humidity systems.[6, 7, 31] When a significant thermal gradient is present in an acoustic field, such as the one between the room temperature and cooling stream, aerosols can be directed toward pressure nodes.[33] In this case, very small water particles in the air would be displaced towards the levitating droplet positions. These particles are easily supercooled, forming INPs when frozen and causing nucleation when they impact cooling liquid droplets. As such, the main factors affecting nucleation time would be the size of the impact area (surface area) and the distance from the cryogun base, the coldest region of the system with the greatest likelihood of INP formation. This could explain why the closest droplet to the cryogun, at position 1, nucleates fastest. The position 2 and position 3 droplets should then nucleate in succession as they become farther from the source of the cooling stream. However, because the position 3 droplet has a larger area for INP impact, nucleation occurs at statistically similar times between these positions. Therefore, the data in this study suggest that secondary nucleation from INPs is the dominant nucleation mechanism at all positions. Finally, it may be further evidence that these parameters (nucleation temperature and time) are not different from those of pure water.[7] Though no thermodynamic effect on crystallization has previously been found for GNF nanofluids, it follows that a secondary mechanism caused by pre-existing particles outside of the droplet would not be affected by the presence of nanoparticles within the droplet.[29, 34] The nanofluid could have affected the nucleation time, specifically reducing it, by enhancing heat transfer out of the liquid droplet. This said, no such effect was measured due to the high driving force for nucleation. An impact could potentially be detected by reducing the driving force. However, this is an intentionally open system, and lower driving forces may be unable to compete with ambient heat from the room so that no crystallization occurs. Closing the system is one solution to examine a single droplet, though these systems lack the positional control necessary to assess multiple droplets of similar volume.

### 3.1.3 Nucleation of Multiple Droplets

The presence of multiple droplets in the system affects nucleation times, causing droplets to freeze simultaneously. When frozen individually, the data in **Table 1** for the droplets in positions 2 and 3 show that they nucleated later than the droplet in position 1. In the 1/2 and 1/2/3 systems, there is no statistical difference between the nucleation times of any droplets, with clear decreases at positions 2 and 3. There may have also been some effect in the 2/3 system, but this was not distinguishable as the nucleation times were already similar. Moreover, there was no effect in the 1/3 system, suggesting that the new multi-droplet effect is related to proximity. Upon forming the primary ice shell, there is an immediate and significant increase in interfacial pressure as the specific volume of ice is greater than that of the water that still comprises most of the droplet core. This pressure likely imposes a sufficiently great mechanical tress to deform and crack the shell.[35, 36] In turn, secondary INPs (i.e., INPs not formed from the water in the air) are emitted by the droplets and then directed toward adjacent droplets by the acoustic field. Furthermore, as mentioned, the cryogenic stream spins the droplets. This improves latent heat transport through forced convection, increasing the initial growth rate, so the instantaneous pressure rises, and the fragmentation frequency is heightened.[11] In brief, secondary INPs released immediately after nucleation may induce nucleation in neighboring droplets in multi-droplet configurations by increasing local INP concentrations. This results in statistically similar nucleation times in those systems and does not apply to the 1/3 system where the droplets are nonadjacent and secondary INPs may be directed towards the empty node at position 2. This behavior was previously observed in levitated crystallization studies using pure water.[7] Therefore, the presence of GNFs did not avert cracking or INP discharge. Instead, it is possible that the GNFs reduced the strength of the primary ice shell, resulting in larger cracks and more INP emission. Some evidence is available in **Figure 2**D, where the deformation of the initial shell was significant enough that the volume of water that escaped could bead at the base of the droplet and was not immediately frozen upon exit. However, this alone is insufficient, as it occurs occasionally in pure water; further discussion will be provided in the next section.

3.2 Solidification of Levitated Graphene Nanofluid Droplets

The post-nucleation stages up to the total solidification of the droplet are shown in **Figure 3** through both morphological and thermal behavior in the 2/3 configuration. Once more, the behavior presented in the figure is characteristic of all droplets in either single or multi-droplet systems. Once the initial shell forms, it grows inward toward the droplet center, as in **Figure 3**A. Sometime later, protrusions would form on the droplet surface, and the droplets would become whiter, losing their initially black appearance. This effect, visible in part B of the figure, was called optical clarity loss (OCL) and results from the combination of lighting, bubble entrapment in the crystal lattice, protrusion formation, and droplet spinning. This moment also marked the continued cooling of the droplet after the heat release of nucleation. As the solid front grew within the droplet, the protrusions outside also expanded, becoming longer and covering more of the surface. Like in **Figure** 3C, protrusions could be fairly thick but converted to more wisp-like forms over time. This difference is evident between the top and bottom droplets in the figure. Though it could occur before this time, after two minutes, the droplet was considered completely crystallized. This meant that (1) there was no liquid remaining in the droplet and (2) heat generation from crystal growth ceased.

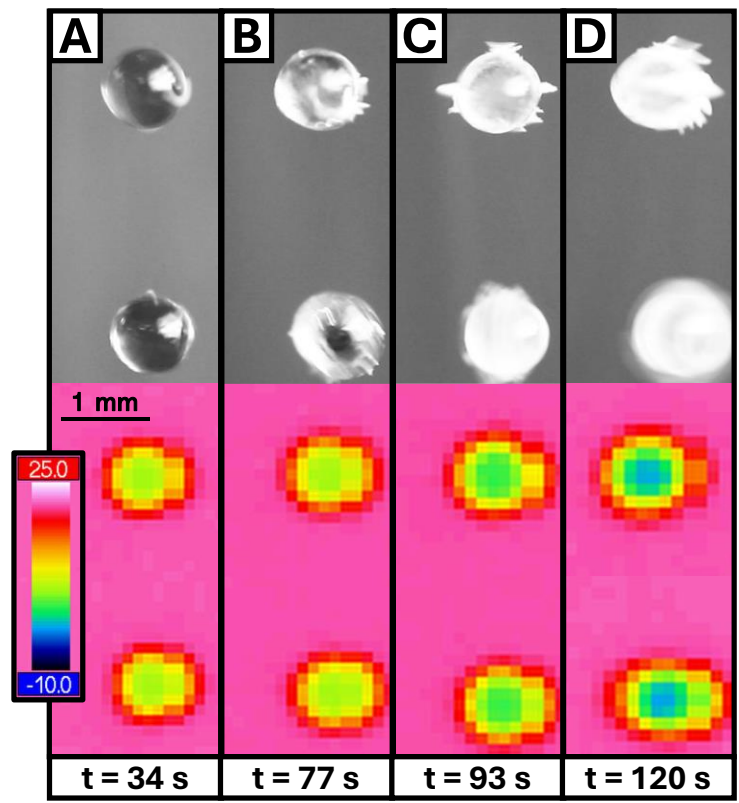

**Figure 3.** A representative example of the morphological and thermal evolution from initial protrusion formation (A) to complete solidification (D) in the 2/3 system. During this process, the droplets lose optical clarity (B), and protrusions grow (C).

### 3.2.1 Bulk Crystallization and Interfacial Protrusion Growth

Starting with the solidification of the liquid bulk, the initially black droplet appears to become white as it freezes. The solubility of air in the crystal phase is many orders of magnitude smaller than that in pure liquid water.[37] As the solid front grows, the dissolved air concentration in the liquid phase increases to a saturation limit at which bubbles are produced. If the growth rate is relatively small, a single large bubble would be present once no liquid remains in the system, as air would have sufficient time to diffuse through the bulk. This occurred in pure water: the droplet remained clear with one bubble (or a small number of larger bubbles) after complete freezing.[7] The crystal growth rate must be significantly greater for many small bubbles to form and become trapped in the solid ice, such that light's path is modified as it passes through the droplet and OCL occurs. It follows that the first time OCL is noticeable, there is an accompanying decrease in temperature. As the solid front grows and more bubbles become trapped in the crystal matrix, the cooling rate from the cryogenic stream eventually overcomes the exothermic heat release. Therefore, OCL and cooling occur on the same time scale.

Critically, this means that despite having the same nucleation times and temperatures in an identical system, levitated GNF nanofluid crystallization occurs faster than levitated pure water crystallization. A previous study examining levitated tetrahydrofuran hydrate (THF) formation in the same system had similar findings. In that study, OCL was also attributed to higher crystal growth rates, which, in that case, were greater because THF hydrates freeze at 4 ºC, so thermodynamic conditions were more favorable.[6] Additionally, previous studies have shown that the number of inhomogeneities in the crystal structure correlates directly with the growth rate.[38] As mentioned, GNFs have been shown to improve mass transfer during liquid-to-solid phase change, which could reduce overall liquid temperature and provide more optimal thermodynamic conditions for continued crystal growth. However, this should also mean that the mass transfer of gases (like air) would also be boosted. Consequently, the effective rise in heat transfer must be greater than that of mass transfer, so more bubble formation still occurs. This may be a system-specific effect for both the cooling rate and the chosen nanoparticle type; it is not clear from the available data that this must be the case, only that the crystal growth rate is higher. In the future, computational studies are recommended to confirm the hypothesis that the liquid heat transfer improvements in nanoparticle-enhanced crystallization systems are greater than those of gas mass transfer. Finally, it can be noted that no multi-droplet OCL effect was detected, likely because the increased number of INPs outside the droplet has no impact on growth rates within.

Beyond bulk crystal growth, several protrusions (like in **Figure 3**C) formed on the frozen surface of the droplet. These protrusions were likely made of ice and seeded by INPs from the air or emitted via interfacial cracking after nucleation.[7] A quasi-liquid layer may be created on the droplet exterior due to pressure changes and heat evolution upon nucleation, resulting in some minor ice shell melting.[33] Discrete INPs could then impact this layer to become the "root" of a protrusion, which would grow through the agglomeration of small ice particles. **Table 2** shows the time required from the placement of the cryogun to the formation of the first protrusions on the droplet. Previous water-based levitation studies have described protrusion formation as stochastic: protrusions could form anywhere from 8 to 35 seconds after nucleation (in the context of this study, an enormous time frame) if they occurred at all, and there were no detectable multi-droplet effects.[6, 7] In this study, protrusion formation was consistent and, examining **Table 2**, there was a clear trend when droplets were adjacent. Specifically, protrusion times decreased in the 1/2 and 2/3 systems and became even shorter (though within error) in the 1/2/3 system. Considering both this and the pure water study had similar gaseous environments and pre-nucleation INP concentrations, a greater and faster occurrence of protrusion formation likely comes from a higher concentration of INPs created by interfacial cracking during the nucleation process. This may mean that there are more and larger cracks in levitated freezing nanofluids than in pure water. This makes sense because, as discussed, the GNFs created an environment for a faster initial growth rate and a higher instantaneous nucleation pressure. Moreover, post-nucleation INP release has previously been correlated to this pressure change.[7] The shell may be further weakened as higher growth rates result in more defects in the crystal lattice, where bubbles and nanoparticles can be trapped in inter-dendritic spaces.[38] Therefore, more INPs are released from the weaker primary ice shell during nanofluid crystallization, and protrusion behavior is constant. As the number of adjacent droplets increases to 2 and then 3, and the number of post-nucleation INPs grows further, the protrusion time decreases by over a third. This result has implications for nanofluid crystallization that should be further explored in future studies, regardless of whether the droplets are levitated or sessile. Namely, suppose an aqueous nanofluid system enhances the crystal growth rate compared to pure water, which could be considered advantageous. In that case, it may have the negative side effect of a weaker solid phase with more lattice defects. This tradeoff may become significant in some phase change technologies, and novel solutions must be found to ensure viability.

**Table 2.** The time for the first protrusions to form on 100 ppm GNF droplets at each single position and multi-droplet configuration. The numbers in brackets following these values represent the 95% confidence interval (±).

| Position | Single     | 1/2        | 1/3        | 2/3        | 1/2/3      |
|----------|------------|------------|------------|------------|------------|
| Pos. 1   | 31.9 (9.5) | 21.2 (5.5) | 28.5 (4.1) | -          | 17.6 (3.1) |
| Pos. 2   | 34.7 (4.9) | 20.3 (5.2) | -          | 23.3 (3.4) | 19.1 (2.8) |
| Pos. 3   | 35.1 (8.0) | -          | 28.0 (7.2) | 23.6 (5.1) | 19.4 (2.7) |

### 3.2.2 Bead Formation and Volumetric Expansion

Different from protrusion growth was the development of liquid beads at the droplet base, such as in **Figure 2**D, which would freeze as the experiment progressed and lower the final droplet sphericity. These beads are always aligned with the vertical axis because liquid flows in the direction of gravity. **Table 3** shows the frequency of bead formation as a percentage of runs where it occurred. Note that there is no uncertainty for these values because they are binary; droplets either did or did not form beads. However, because they are out of 15, a change in the behavior of one droplet has a 7% effect on the value, so they should only be taken roughly. From the table, around 50 to 60% of droplets formed beads at position 1 and position 2, whereas the formation frequency was consistently 27% at position 3. Droplet adjacency did not affect these values, which suggests that INPs do not play a role in their formation, and the cause lies inside the droplet. Indeed, beads likely form when bulk water, pushed by high internal pressures, escapes the droplet through pores and defects in the ice shell.[32] Therefore, bead formation is a product of the initial growth rate, which governs the instantaneous pressure at nucleation. It has previously been suggested that early protrusion formation has the greatest effect on sphericity for freezing levitated water.[7] However, if this were the case, the 1/2/3 system would have had more bead formation, making it more likely to be a mechanical effect. More precisely, while protrusions may have some effect, initial pressure effects are dominant in systems with lower primary shell integrity. This is bolstered by findings in the THF hydrate levitation study, which attributed the greater mechanical strength of THF hydrate compared to ice to the absence of beading behavior.[6] It may be that the position 3 droplet, which is furthest from the cryogun and possibly has the lowest internal cooling rate due to its larger size, has the smallest initial growth rate compared to the positions above, resulting in lower instances of bead formation. This potential effect also applies to the observation that while 27% was the smallest formation frequency in this study, it was the largest for pure water (whose lowest was 6%).[7] The higher growth rates in levitated freezing nanofluids result in more defects and cracks in the initial shell, with greater internal pressures, so the occurrence of beading is also increased.

**Table 3.** Bead formation and volumetric expansion when freezing 100 ppm GNF droplets at each single position and multi-droplet configuration. The numbers in brackets following these values represent the 95% confidence interval (±).

| Configuration | Single | 1/2 | 1/3 | 2/3 | 1/2/3 |
|---|---|---|---|---|---|
| Bead Formation (%) | | | | | |
| Pos. 1 | 60 | 53 | 67 | - | 53 |
| Pos. 2 | 53 | 47 | - | 53 | 53 |
| Pos. 3 | 27 | - | 27 | 27 | 27 |
| Volumetric Expansion (%) | | | | | |
| Pos. 1 | 26 (10) | 27 (15) | 27 (8) | - | 28 (9) |
| Pos. 2 | 24 (5) | 29 (8) | - | 29 (9) | 31 (9) |
| Pos. 3 | 25 (7) | - | 31 (6) | 21 (9) | 35 (11) |

The droplets were frozen solid once the two-minute experimental time had elapsed, and no liquid remained in the bulk. The final droplets looked like those in **Figure 3**D, with a frozen white core containing entrapped bubbles and nanoparticles, some protrusions, and either a spherical or beaded sphere shape, depending on whether a bead formed. This time also marked the droplet's coldest point, as no more heat generation from crystal growth was present. The expected volumetric expansion for water during crystallization is about 9%.[39] The volumetric expansion observed in this study is presented in **Table 3** with rounded values due to the high uncertainty of length measurements when the cryogenic stream jostles the droplets. Note that this measurement does not include protrusions, only the solid core, though it did include the bead. Values are much higher than expected, ranging from 25 to 30% and approximately 28% on average. Volumetric expansion was not influenced by droplet position or adjacency. The ideal 9% expansion considers a defect-free crystal lattice structure made from pure, sessile water. The entrapped air that is characteristic of levitated nanofluid crystallization, as well as grain boundaries from high growth rates, may significantly expand this lattice. Moreover, the presence of beads leads to a higher average final droplet diameter: levitated THF hydrate freezing, which does not form beads, saw only a 24% expansion.[6] The acoustic field may also influence crystal growth kinetics to introduce morphological defects. Oscillations at the phase boundary between the solid crystal front and aqueous solution could modify the crystal structure on a large scale, lowering the effective density of the droplet and augmenting volumetric expansion in the present timescale.[40] Cracking from initial internal pressures is unlikely to contribute to volumetric expansion because cracks are mostly in the length scale of microns, while expansion is millimetric. If cracking were the same magnitude as the droplet, it would likely shatter catastrophically. As such, pure water had similar volumetric expansion effects with an average of 31%.[7] This makes sense as GNFs are not expected to alter ice crystal structure substantially. Previous studies have shown that GNFs do not impose an additional expansion effect as they have high mobility at the solid front.[38] Due to their size, shape, and hydrophilicity, GNFs become entrapped in bubbles and inter-dendritic spaces that are already solid (i.e., ice does not form around GNFs to immobilize them; they slip into open spaces that are already part of the crystal structure). Therefore, despite the higher growth rate effect in nanofluids resulting in more small bubbles and defects, the overall impact of these changes may be within error and, thus, undetectable.

3.3 Melting of Levitated THF Hydrates

After the two-minute crystallization time, the cryogun was removed, and the solid droplets were again exposed to room-temperature air. The melting behavior that followed over the next minute was observed at all positions and can be seen in **Figure 4**, using an example from the position 2 system. At the first moment of melting, like in **Figure 4**A, the droplet still has protrusions and is spinning from the force of the cryogenic stream. Soon after, the protrusions vanish, likely melting quickly due to their low density. The droplet stops spinning and melts, becoming more oblate as it liquefies around the ice core (B). As crystal decomposition progresses, intra-droplet convection currents form, and a flow of water, ice, and nanoparticles develops about the lateral axis (C). Finally, no solid ice remains, flow ceases, and the droplet returns to its initial temperature (D). The final droplet contains large air bubbles and has a lighter color compared to its pre-crystallization condition (see **Figure 2**A, which is also in the position 2 system).

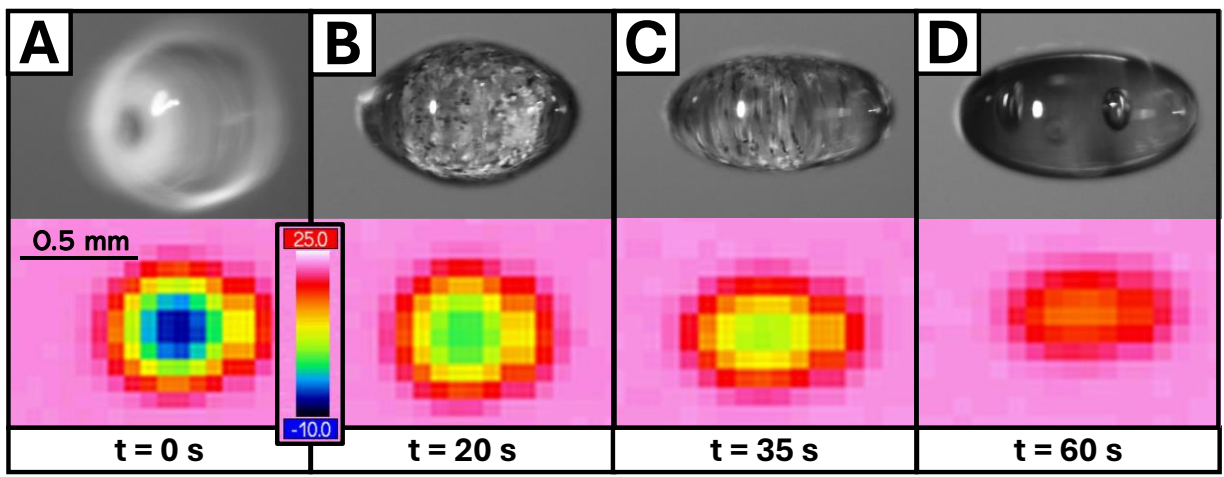

**Figure 4.** An example of the morphological and thermal evolution from complete crystallization (A) to reaching the initial temperature (D) at position 2. While the droplets thaw, entrapped

nanoparticles and bubbles are visible (B). Eventually, convective currents form, and the droplet spins about the horizontal axis (C).

### 3.3.1 Convective Current Formation and Final Droplet Conditions

The flow effect in melting levitated droplets has been reported previously and shares similarities with melting behavior in sessile nanofluid droplets.[6, 7, 41] In either case, after some melting occurs on the outside of the droplet, the solid core does not remain completely centered but rises due to buoyancy. This creates a thermal gradient with the top of the droplet closer to the ice temperature and the bottom closer to room temperature. A surface tension gradient forms vertically across the droplets, producing convective cells about the horizontal axis. The currents initially form underneath the ice, starting at the droplet center and moving downward, but eventually overcome the buoyant force and take up the remaining ice.[41] When thermal gradients result in surface tension-driven flow, it is called Bénard-Marangoni convection, and such systems will tend toward high Marangoni numbers (Ma). In pure water, the value of Ma was calculated to be $1.9 \times 10^4$, which is sufficient to indicate that the thermal gradient generates convective cells.[7] In this study, however, the value is likely lower because GNF droplets have slightly lower surface tension values and better transfer properties, leading to a thermal gradient that abates more quickly. The latter is evident from **Figure 4**D, where the thermocapillary effect is no longer present after one minute, whereas intra-droplet flows persisted beyond one minute in pure water.[6, 7] An exact value for Ma cannot be calculated because the droplets are poorly suspended after a single freeze-thaw cycle, and the effective properties necessary for calculation, particularly while the system undergoes flow behavior with much greater length scales than the nanoparticles, cannot be measured. The value is at least $10^3$, the minimum order of magnitude necessary for the observed convective flows, and likely closer to $10^4$, the order for pure water.

Melting was complete after one minute had elapsed under room temperature, even if some droplets returned to their initial temperature a few seconds before. This ensured that the same observation period was maintained for every droplet. Regardless of configuration, the final volume did not change significantly from the initial one: across all experiments, the average change in volume was 0.07 ± 0.08 µL. Therefore, there was no statistical difference between the initial and final volumes, consistent with other levitated crystallization studies.[6, 7] Volume changes are not expected during the levitated crystallization-decomposition cycle. While mass may be added from atmospheric water forming protrusions or lost due to acoustic streaming (evaporation) and INP production, these would likely only produce changes within error. Previous estimates in similar systems have proposed volumetric losses of 0.5% from evaporation over one minute.[42] However, if we consider the largest droplets with the maximum evaporation surface area, the loss would be 0.015 µL, which is not significantly high. Therefore, while there may be effects that add and subtract volume from the droplets over an experimental run, they are negligible, and the liquid volume is statistically constant.

3.3.2 Relative Suspension Loss

When comparing the darkness of the droplet before and after the freeze-thaw cycle, there is a noticeable shift from a fairly black droplet to a gray one. This difference is shown most clearly in **Figure 5**C between the initial and post-melt droplets. Significant losses in suspension stability are known to occur in GNF nanofluids during freezing.[38] Such a loss in this study would make the droplets more translucent (i.e., reduced opacity) and reflective (i.e., reduced light absorbance). This occurs because GNF clusters form during solidification, most visible in **Figure 4**B-C, and do not spontaneously redisperse after melting. It may be expected that plasma-functionalized (hydrophilic) nanoparticles return to their initial dispersion when the crystal matrix is completely removed. However, physical and electrostatic effects during crystallization modify the suspension properties to where an additional energetic input (e.g., sonication) is required to return to the initial state. After the primary shell formation, and as more liquid water becomes incorporated into the solid ice, a drying process occurs where the concentration of GNFs in the bulk increases. This increases the collision probability, further enhanced by the high intra-droplet pressure in the liquid as the surrounding ice expands. Together, these effects shift the repulsive potential between the GNFs into a more attractive regime, and they come together.[43] Once together, energy is required to overcome van der Waals forces and separate individual particles because the interparticle repulsion mechanism in the initial droplet is less present. Specifically, the electrostatic double layer between particles, which requires a surrounding liquid solution to dissolve charged ions, is absent. When the base fluid is removed between the GNFs during freezing, there is no medium for ion exchange and no repulsion.[43] During melting, it is highly unlikely that water will spontaneously exert pressure to force itself between two nanoparticles in contact, so the double layer does not reform and the particles stay together.

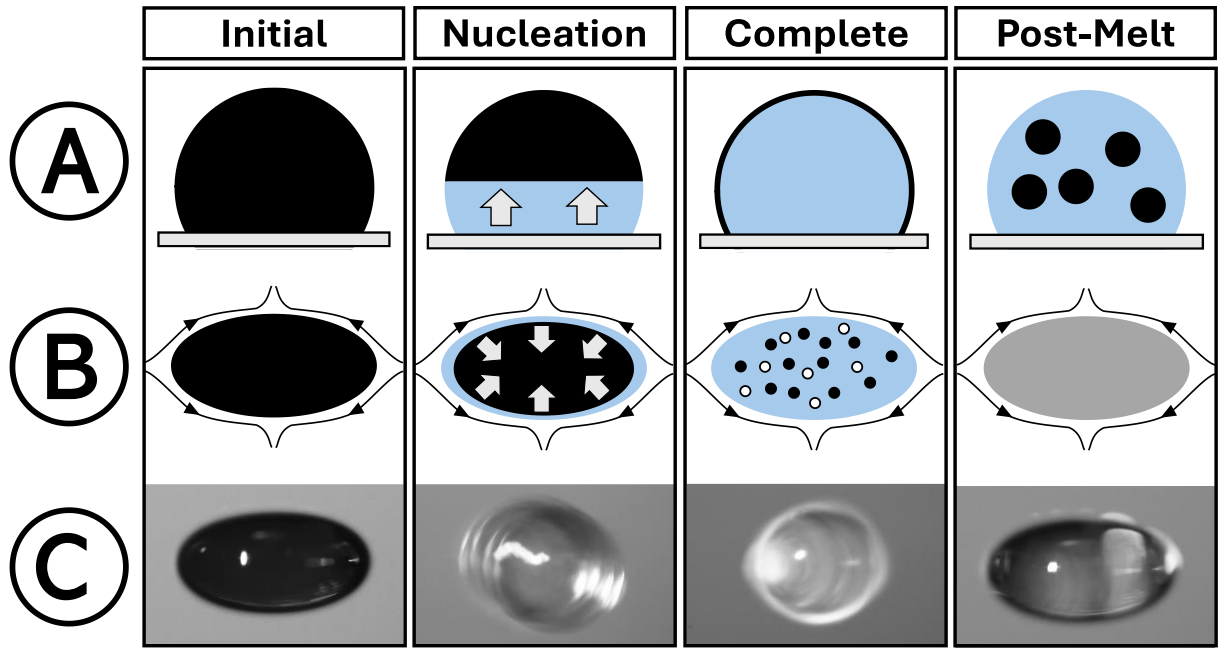

**Figure 5.** The loss of nanofluid suspension stability during a freeze-thaw cycle. Cross sections are provided for droplets at rest on a solid surface (A) and floating in the levitator (B). A visual example at position 2 for effects observed in this study is provided in (C).

It is difficult to quantify the exact amount of dispersion loss without touching the droplets, and the physical handling of such small samples would significantly affect the result. Therefore, the mean gray value (MGV) was used as a contact-free estimate of the suspension loss. This unitless value is the average "distance" between black (0) and white (255) of the pixels making up the image of a droplet. The initial MGV of the droplets was $32.7 \pm 6.6$ and $81.9 \pm 7.2$ for the single and multi-droplet configurations, respectively. These values are much closer to 0 than 255, indicating fairly black droplets. The multi-droplet configurations have a higher base value because the camera is less zoomed in (so that all droplet positions are visible) and can take in more light from the surroundings, making the droplets appear lighter. The final MGVs were $84.7 \pm 12.2$ and $142.9 \pm 9.1$, suggesting much lighter droplets with worse suspensions. The droplets may also be less homogenous as bubbles could contribute some glare and increase the final MGV. This said, the values on their own have no physical significance. Instead, the change in these values may represent a relative suspension loss. **Equation 1** was used to calculate the value of these changes. However, it was necessary to normalize the resultant values to the MGV of the base fluid because the maximum possible MGV in the system is not 255, but the MGV of pure water.

$$\Delta MGV_{norm} = \frac{MGV_f - MGV_i}{MGV_{bf} - MGV_i} \qquad \textbf{(Equation 1)}$$

Using $MGV_{bf}$ values of 116.2 and 167.7, the water-normalized changes in MGV were $0.62 \pm 0.04$ and $0.75 \pm 0.05$ for single and multi-droplet systems, respectively. Therefore, only considering color, GNF droplets became 62% closer in appearance to pure water in the single system (as the change accounts for 62% of the maximum possible change) and 75% closer in the multi-droplet systems. This suggests that the dispersion in multi-droplet systems is worse after one crystallization cycle than when droplets are frozen alone. Previous studies have shown that lower crystal growth rates result in greater dispersion loss, so this result may suggest that when multiple droplets freeze simultaneously, their growth rates are lower.[38] However, no additional evidence suggests this was the case or that multi-droplet effects played a role in bulk crystallization. Instead, it may be that the same aperture effect that increases MGV values when zooming out also exacerbates changes in MGV. Therefore, what can only be concluded based on these values is that the dispersion loss results in a decrease in light-related effects (i.e., lower transmission and absorbance) by at least 62%.

Finally, compared to images after a freeze-thaw cycle in sessile GNF droplets (i.e., those frozen on solid surfaces), the dispersion of levitated GNFs appears superior. Previous sessile droplet studies have shown that freezing 100 ppm GNF nanofluids at the temperatures used in this study should result in the particles being wholly expelled upon freezing and forming large clusters after melting.[38] This is depicted in **Figure 5**A. In this study, the GNFs cannot be expelled from the solid matrix because of the primary ice shell and display much smaller clusters (**Figure 5**B). The ice shell likely plays the most significant role in this difference because of particle entrapment and higher pressures. For instance, because the particles cannot leave the droplet, they are trapped in defects in the crystal matrix rather than as a mass outside it. This means they are more separate from each other before melting. Similarly, air is trapped in the droplet, so bubbles form as another GNF entrapment space. In sessile droplets, there is sufficient time and surface area exposed to the room for air to completely escape (at the temperatures observed in this study). Moreover, higher pressures within the droplet result in cracks and inter-dendritic spaces in the ice where GNFs can remain when no more liquid is present. These result from losing the isobaric nature of the sessile process. Moreover, because the cryogenic stream surrounds the entire droplet and there is no single solid cooling surface, the average levitated droplet temperature is likely cooler, and growth rates are higher. Higher growth rates would induce the same pressure-related defects as the presence of the primary ice shell. Both sessile and levitated droplets also produce convective currents when melting, though these are stronger in levitated droplets, which can also play a dispersive role.[6, 7] Therefore, levitated nanofluid studies could reveal the effects of non-isobaric crystallization where the cooling mechanism surrounds the entire sample: greater pressures in an enclosed environment produce more defects that, in turn, result in a better final dispersion, requiring less energy to resuspend.

CONCLUSIONS

Aqueous droplets containing a 100 ppm concentration of plasma-functionalized GNFs were suspended in an acoustic levitator and underwent a crystallization-decomposition cycle.

Selecting different nodal positions, IR and digital cameras simultaneously observed configurations of up to three droplets during nucleation, crystal growth, and melting. Like in many levitated crystallization studies, crystal growth was initiated by ice nucleation particles (INPs) formed from atmospheric water. Due to pressure-induced cracking of the primary ice shell that develops around the droplets upon nucleation, more INPs were released into the acoustic field, which meant that droplets that nucleated at separate times alone now nucleated simultaneously. As this effect generally comes from factors outside the droplet, the GNFs within the liquid bulk did not play a role such that nucleation times and temperatures were not statistically different from those of pure water. However, GNFs are crystal growth rate enhancers via improved mass transfer phenomena, so the initial growth rates (and, thus, the instantaneous pressure at nucleation) are greater than in pure water, leading to more significant cracking behavior and INP production. In turn, protrusions appear much more rapidly on the surface, and liquid water more readily escapes through defects to develop into beads at the droplet base. Moreover, higher growth rates induce defects in the crystal matrix, including inter-dendritic spaces and entrapped bubbles. As phase change proceeds, these modify the optical properties of the droplets such that they appear white rather than black (as they were initially) or clear (as pure water). Volumetric expansion in the droplets was about 28%, much higher than the theoretical 9%, though similar to the expansion of levitated pure water. In this case, GNFs did not affect volumetric expansion because the new defects they induce are a full order of magnitude smaller than the droplet diameter.

During crystal decomposition, thermal gradients develop vertically across the droplet, creating a surface tension difference that initiates convective flow. The flow in GNF droplets was likely weaker than that in pure water, as the same effects that enhanced crystal growth would also enhance crystal decomposition, leading to warmer average temperatures while the droplet melted.

As such, the droplet was fully melted after one minute, and the convective cells had dissipated, whereas liquid flow had not ended by this time in pure water. Finally, all droplets experienced a significant loss in suspension stability after the freeze-thaw cycle. This was due to GNFs becoming trapped in bubbles and interdendritic spaces during freezing or coming together due to high pressures and collision frequencies within the liquid bulk. There was no water between the GNFs for ion exchange after melting, so they did not spontaneously redisperse. The decrease in light-related effects, like droplet opacity and absorbance, from particle clustering made the droplets appear at least 62% more similar to pure water based on mean gray value analysis. However, the levitated dispersion had smaller clusters than when sessile GNF droplets were frozen on solid surfaces at the same temperature. Due to the formation of a primary ice shell, levitated crystallization is non-isobaric, so more bubbles form and particles cannot be expelled from the crystal matrix. In addition, growth rates are higher at the same temperature because the cryogenic stream surrounds the droplet, so there are more defects in which the GNFs can be trapped. As such, levitated nanofluids have a better dispersion, mainly because less energy would be required for nanoparticle resuspension from smaller clusters.

## AUTHOR INFORMATION

**Corresponding Author**

*phillip.servio@mcgill.ca

**Author Contributions**

The manuscript was written through the contributions of all authors. All authors have approved the final version of the manuscript.

ACKNOWLEDGEMENTS

The authors acknowledge the financial support from the Natural Sciences and Engineering Research Council of Canada (NSERC) and the Faculty of Engineering of McGill University (MEDA, Vadasz Scholars Program). We also thank Anne-Marie Kietzig and the Biomimetic Surface Engineering Lab at McGill for the use of their TinyLev device.

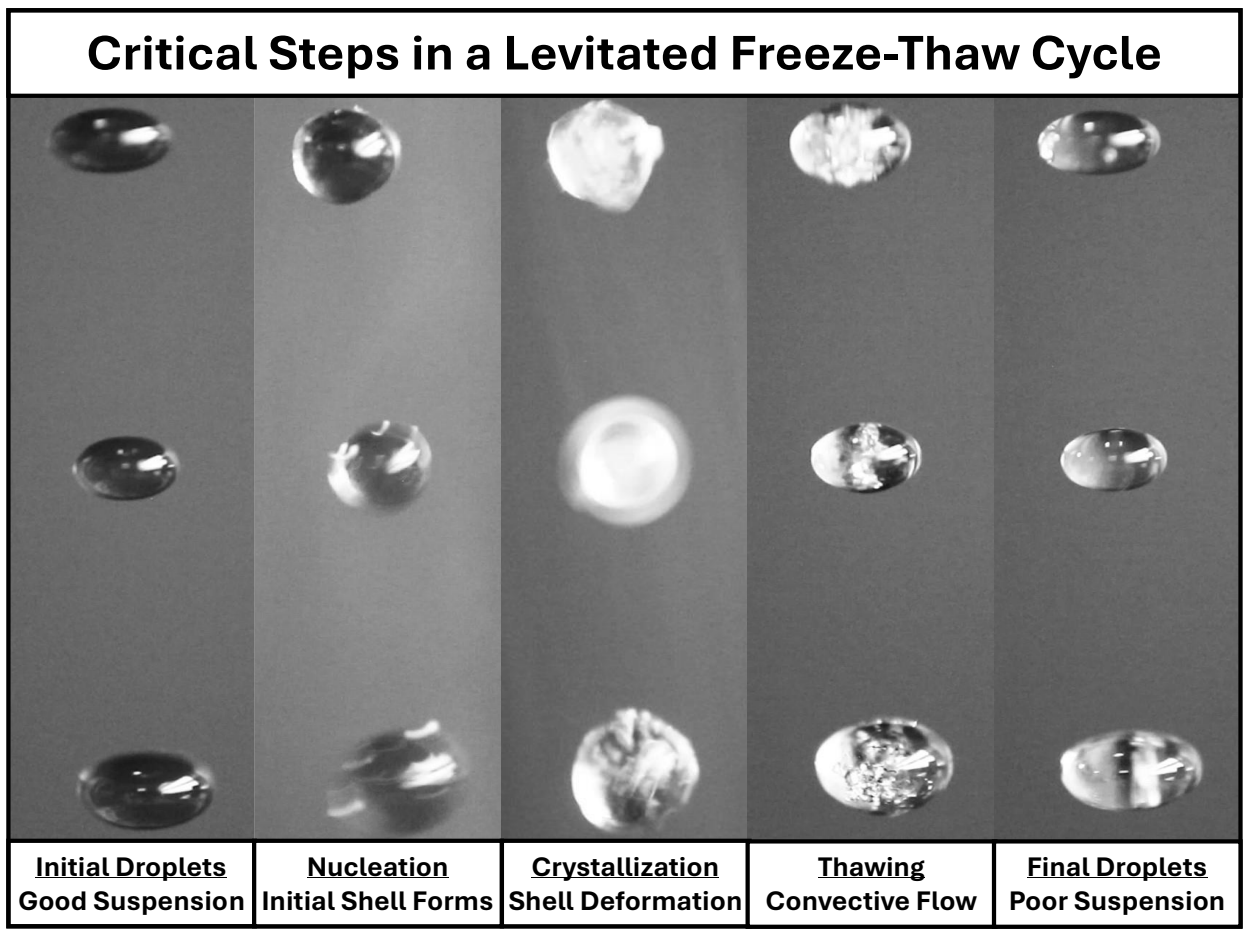

Graphical abstract for review purposes only